\documentclass[runningheads]{llncs}
\usepackage[T1]{fontenc}
\usepackage{placeins}
\usepackage{amsfonts}
\usepackage{amsmath}
\usepackage{cleveref}
\usepackage{graphicx}
\usepackage{color}
\usepackage{array}
\usepackage{cite}
\usepackage{graphicx}
\usepackage[table,xcdraw]{xcolor}
\usepackage{multicol}
\usepackage{adjustbox}
\usepackage{colortbl} % For coloring cells
\usepackage{booktabs}
\usepackage{multirow}
\usepackage{caption} 
\usepackage{subcaption} 
\usepackage{float}
\usepackage{enumitem}
\setlist{nolistsep}

\newcommand{\boldgreen}[1]{\textcolor{green!60!black}{\textbf{#1}}}
\newcommand{\boldred}[1]{\textcolor{red!80!black}{\textbf{#1}}}

\newcommand{\ecl}{\textsc{Eclipse}}

\newcommand{\bfs}{\mathbf{s}}
\newcommand{\bfd}{\mathbf{d}}
\newcommand{\bfa}{\mathbf{a}}
\newcommand{\bfp}{\mathbf{p}}
\newcommand{\bfq}{\mathbf{q}}
\newcommand{\bfm}{\mathbf{m}}
\newcommand{\supa}{\textsuperscript{\textit{a}}}
\newcommand{\supb}{\textsuperscript{\textit{b}}}
\newcommand{\supab}{\textsuperscript{\textit{ab}}}

\newcommand{\sstar}{\textsuperscript{*}}

\newcommand{\midsepremove}{\aboverulesep = 0.1mm \belowrulesep = 0.3mm}
\midsepremove
\newcommand{\midsepdefault}{\aboverulesep = 0.605mm \belowrulesep = 0.984mm}
\midsepdefault

\begin{document}
\title{\ecl: Contrastive Dimension Importance Estimation with Pseudo-Irrelevance Feedback\\ for Dense Retrieval}
\titlerunning{\ecl: Contrastive DIME with Pseudo-Irrelevance Feedback}
\author{Giulio D'Erasmo\inst{1} \and
Giovanni Trappolini\inst{1} \and
Nicola Tonellotto\inst{2} \and \\
Fabrizio Silvestri\inst{1}}
\authorrunning{D'Erasmo et al.}
\institute{
Sapienza University of Rome, Italy \\
\email{\{g.derasmo, trappolini, fsilvestri\}@diag.uniroma1.it} 
\and
University of Pisa, Italy\\
\email{nicola.tonellotto@unipi.it} 
}
\maketitle              % typeset the header of the contribution
\begin{abstract}
Recent advances in Information Retrieval have leveraged high-dimensional embedding spaces to improve the retrieval of relevant documents. 
Moreover, the Manifold Clustering Hypothesis suggests that despite these high-dimensional representations, documents relevant to a query reside on a lower-dimensional, query-dependent manifold. 
While this hypothesis has inspired new retrieval methods, existing approaches still face challenges in effectively separating non-relevant information from relevant signals. 
We propose a novel methodology that addresses these limitations by leveraging information from both relevant and non-relevant documents. 
Our method, \ecl, computes a centroid based on irrelevant documents as a reference to estimate noisy dimensions present in relevant ones, enhancing retrieval performance. 
Extensive experiments on three in-domain and one out-of-domain benchmarks demonstrate an average improvement of up to $19.50\%$ (resp. $22.35\%$) in mAP(AP) and $11.42\%$ (resp. $13.10\%$) in nDCG@10 w.r.t. the DIME-based baseline (resp. the baseline using all dimensions).
Our results pave the way for more robust, pseudo-irrelevance-based retrieval systems in future IR research.

\end{abstract}

% introduzione
\section{Introduction}

Dense retrieval models \cite{karpukhin-etal-2020-dense, devlin-etal-2019-bert, colbert} have revolutionized Information Retrieval (IR) by embedding both queries and documents into a latent, high-dimensional vector space, where their similarity can be computed more effectively \cite{approximatenearestneighbor, IR_system}. In this latent space, the dimensions of the embeddings represent different features that the model has learned to be important for capturing the text's meaning or other latent characteristics of the encoded text \cite{liu-etal-2022-dimension, nguyen2016neuralbasednoisefilteringword, yin2018dimensionalitywordembedding}.

However, not all dimensions in this space contribute equally to retrieval performance, and some may add noise or unnecessary information \cite{lower-dimensional, al-sharou-etal-2021-towards}. This creates a need to focus on the most important dimensions, similar to how feature selection works in machine learning \cite{featureselection, featureselection2}.

To address this, researchers have proposed various dimension reduction techniques \cite{ba2019discoveringtopicsneuraltopic, efficientqa, ma-etal-2021-simple}, including the Manifold Clustering Hypothesis by Faggioli et al.~\cite{DIME}, which suggests that queries and documents lie on lower-dimensional subspaces that are specific to each query, where only the most relevant dimensions are retained.
A practical implementation of this idea is offered by tools like Dimension IMportance Estimators (DIMEs), which rank dimensions in the latent space by their importance and retain only the most relevant ones. DIMEs compute importance scores for each dimension based on a \textit{relevant feedback document}, using methods such as Pseudo-Relevance Feedback (PRF) \cite{rocchio1971relevance, prf2} and Large Language Models (LLM) \cite{devlin-etal-2019-bert, reimers2019sentence, izacard2021unsupervised, radford2018improving}. In the PRF-based DIME approach, the embeddings of the top-ranked documents—retrieved from the corpus using similarity measures like cosine similarity or the inner product—are averaged to form the pseudo-relevant document representation. Similarly, in the LLM-based DIME, an LLM generates a document (e.g., in response to the query) that acts as the relevant feedback document.

Despite their effectiveness, current DIME approaches face challenges distinguishing between relevant and irrelevant dimensions, potentially retaining noisy dimensions that can reduce retrieval performance. This occurs because there is a need for balance between the original query and the feedback information, as over-trusting this feedback can lead to a bias toward a particular subset of documents while under-trusting it may prevent us from fully leveraging the feedback's potential \cite{adaptiverelevancefeedback, datta2024deep, rawte2023survey}. These challenges stem from the fact that DIMEs focus solely on relevant documents. We hypothesize that irrelevant documents can provide a powerful tool, helping to more clearly delineate the truly important dimensions, like in contrastive learning, where both positive and negative examples are crucial for identifying discriminative features \cite{chen2020simpleframeworkcontrastivelearning, goodfellow2015explainingharnessingadversarialexamples, mikolov2013efficientestimationwordrepresentations}.

Building on this hypothesis, we introduce \ecl, a novel method that leverages relevant and irrelevant document representations to more accurately identify important dimensions. \ecl\ constructs a representation of a generic \textit{pseudo-irrelevant} document by calculating a centroid of embeddings from retrieved non-relevant documents. This is then subtracted from the relevant document representation, effectively suppressing non-relevant dimensions. \footnote{The name "\ecl" draws an analogy to an annular solar \ecl, where the Moon covers most of the Sun, leaving only its outer edge visible – similarly, our method "eclipses" irrelevant dimensions to highlight the most important ones.}

With our experiments we aim to address the following research questions:
\begin{itemize}
    \item \textbf{RQ1}: Can pseudo-irrelevance feedback be leveraged to improve state-of-the-art DIME approaches?
    %\modificato{Does our proposed approach significantly improve over standard DIMEs?} 
    
    \item \textbf{RQ2}: What type of irrelevant documents are used to highlight relevant dimensions in the embedding space?
    %\modificato{Does our method actually depend on irrelevant documents to highlight relevant dimensions?}
    
    \item \textbf{RQ3}: Is our approach dependent on the semantic content of irrelevant documents to improve IR system performance?
    %\modificato{Is \ecl\ dependent on the semantic content of the irrelevant documents?} 
\end{itemize}

\textbf{Our contribution:} We introduce \ecl, a novel method that enhances document retrieval by leveraging relevant and irrelevant document representations to more accurately identify important dimensions. \ecl\ can be applied to any standard Dimension IMportance Estimator, improving performance without requiring changes to the underlying retrieval model. This flexibility makes our method a valuable addition to existing IR systems and frameworks. Through extensive experiments with state-of-the-art TREC collections (Deep Learning 2019 \cite{trec19}, 2020 \cite{trec20}, DL-HARD 2021 \cite{trechd}, and Robust 2004 \cite{robust04}), we demonstrate an average improvement of up to $19.50\%$ (resp. $22.35\%$) in AP, and $11.42\%$ (resp. $13.10\%$) in nDCG@10 w.r.t the DIME-based baseline (resp. the baseline using all dimensions).

The rest of the paper is organized as follows. In Section \ref{background}, we provide a comprehensive overview of the DIMEs. We introduce \ecl\ in Section \ref{method-sec}, and we evaluate it against DIME baselines on experiments defined in Section \ref{exp-setting}. Results can be found in Section \ref{results}. Finally, Section \ref{conclusion} concludes our work.

% related works
\section{Background}\label{background}

In dense IR, the dimensions of the textual embeddings do not contribute equally to relevance. Some dimensions introduce noise, reducing retrieval performance. One key approach to retain only a small set of dimensions is the Manifold Clustering (MC) hypothesis~\cite{DIME} which posits that high-dimensional representations of queries and their relevant documents reside within a lower-dimensional, query-dependent subspace than the original high-dimensional latent space.

Let $\bfq \in \mathbb{R}^d$ represent a query and $\mathcal{D} \subset \mathbb{R}^d$ denote a set of documents, where both queries and documents are embedded into a shared latent space $ \mathbb{R}^d $ via a bi-encoder model. This model optimizes query-document alignment by mapping them into a latent space where the query-document relevance is proportional to similarity measures such as cosine similarity or inner product. %The retrieval task is to rank documents by sorting them in descending order of similarity scores. 
The MC hypothesis suggests that the retrieval performance of a dense IR system can be enhanced by projecting both the query $\bfq$ and documents $\mathcal{D}$ into a query-dependent subspace, where irrelevant dimensions are discarded.

DIMEs efficiently assign importance scores to dimensions in $\mathbb{R}^d$ using a query-dependent function $ u_q: \{1, \ldots, d \} \to \mathbb{R} $, which outputs a score for each dimension $ i \in \{1, \ldots, d \} $ in the latent representation for a query $\bfq$. The function $ u_q(i) $ estimates how much the dimension $ i $ contributes to retrieving relevant documents for the given query $q$. This score allows the system to rank the dimensions, keeping those with higher scores and discarding less important ones. The selected dimensions form a low-dimensional query-dependent subspace of $\mathbb{R}^d$.

Two methods for estimating dimension importance are PRF and LLM DIMEs.
 
The PRF DIME utilizes pseudo-relevance feedback by assuming that the top $ k $ documents retrieved by a similarity measure such as BM25~\cite{bm25} are likely relevant to the query. These documents are combined into a centroid vector $\bfp$. Thus, the importance of each dimension $ i $, denoted $u^{PRF}_q@k(i)$, is computed as:

\begin{equation}
    u^{PRF}_q@k(i) = \bfq_i \cdot \bfp_i.
\end{equation}

This importance value captures the alignment between the query $\bfq$ and the pseudo-relevant documents coalesced into $\bfp$, helping to rank and select the most relevant dimensions.

The LLM DIME uses a synthetic document $\bfa$, generated by an LLM, assumed to be relevant to the query. In this case, the importance of each dimension $i$, denoted $u^{LLM}_q(i)$, is calculated as:

\begin{equation}
    u^{LLM}_q(i) = \bfq_i \cdot \bfa_i.
\end{equation}

While able to boost the retrieval effectiveness of dense IR systems in many different scenarios, PRF and LLM DIMEs do not incorporate any further relevance signal provided by other sources, such as partially relevant and irrelevant documents. Relying only on (pseudo-)relevant documents may introduce a noise component into the method, having the document's embedding capture both the relevant and non-relevant parts of the full text. This can be worsened by aggregating multiple texts into a single representative embedding, increasing noise.

\section{Proposed Method}\label{method-sec}

We postulate that the limitations we discussed in the DIME can be overcome by incorporating both relevant and irrelevant documents into the dimension importance function, offering a new direction for refining the retrieval accuracy.

To explore our assumption, we present \ecl, a novel framework aimed at improving dense vector retrieval by incorporating non-relevant documents into the retrieval process. 
In \ecl, for a given query $\bfq$, embedded into a latent space using a bi-encoder, we retrieve a set of $k$ documents ranked according to a relevance score derived from a similarity function. This set of documents, denoted as $\mathrm{D}_q  = \{ \bfd_1, \bfd_2, \dots, \bfd_k\}$, contains pseudo-relevant documents, whose content captures mainly relevant information, and typically found at the top positions, and potentially pseudo-irrelevant documents at the bottom positions, whose content captures mainly irrelevant information. Fixing a parameter $0 < k^{-} < k$, we define a mean vector $\bfm$, named \textit{moon}, as an irrelevant representative embedding, aggregating the embeddings of the bottom $k^-$ documents in $\mathrm{D}_q$, as follows: 

\begin{equation}\label{IRR-mean vector}
    \bfm = \frac{1}{k^{-}} \sum_{i=0}^{k^- -1} \bfd_{k-i}.
\end{equation}

We define our \ecl\, denoted $u^{\text{\ecl}}_q(i)$, as a weighted difference of a relevant representative embedding $\bfs$, the \textit{sun} vector, and the irrelevant representative embedding $\bfm$:

\begin{equation}\label{ecl}
    u^{\text{\ecl}}_q(i) = \alpha (\bfq_i \cdot \bfs_i) - \beta (\bfq_i \cdot \bfm_i).
\end{equation}

The parameters $\alpha, \beta \in \mathbb{R}$ control the balance between the relevant and irrelevant document signals. Instead of using a convex combination, we apply independent weighting to each term, which provides more flexibility and shows superior performance in our experiments.

In Eq.~\eqref{ecl}, the embedding $\bfs$ depends on the original DIME used to compute the relevant signal, defined as: 

\begin{equation*}
    \bfs = 
\begin{cases}\label{cases}
    \frac{1}{k_{+}} \sum_{i=1}^{k^{+}} \bfd_{i}  & \text{if using PRF;}\\
    \bfa,              & \text{if using LLM}.
\end{cases}
\end{equation*}

\noindent
For the PRF case, the term $ \frac{1}{k_{+}} \sum_{i=1}^{k^{+}} \bfd_{i} $ represents the mean vector of the embeddings of the top $0 < k^{+} < k-k^-$ embeddings from $\mathrm{D}_q$. 
For the LLM case, the term $ \bfa $ refers to the embedding of the document generated by the LLM as a response to the query $ q $. 

The interplay between relevant and irrelevant information is a fundamental challenge in optimizing retrieval systems. Indeed, using the Hadamard product, Eq.~\eqref{ecl} can be rewritten as:
\begin{equation}
   u^{\text{\ecl}}_q = \alpha (\bfq \odot \bfs) - \beta (\bfq \odot \bfm) = \bfq \odot (\alpha \bfs - \beta \bfm),
\end{equation}
This formulation highlights the \textit{residual} \ecl\ vector $\alpha\bfs - \beta\bfm$, showing how the irrelevant centroid $\bfm$ suppresses the non-relevant dimensions from the relevant signal $\bfs$. The residual vector is further adjusted by the Hadamard product with the query embedding $\bfq$, which highlights query-relevant dimensions. The more closely the residual \ecl\ vector is aligned with the query, the more it amplifies the contrast between relevant and irrelevant documents.

\ecl\ suppression of irrelevant dimensions, allows the retrieval system to position documents sharing the query's topic higher in the ranking.
Figure \ref{topic-separation} illustrates the effectiveness of our proposed technique in maintaining topic relevance with respect to the query. 
We analyze the top retrieved passages for the query "What is an active margin?" using the model ANCE over the MS-MARCO collection \cite{bajaj2018msmarcohumangenerated}. PRF \ecl\  demonstrates superior performance in retrieving topically relevant documents compared to the PRF Standard DIME, achieving a significantly higher AP (0.498 vs 0.183). While PRF Standard DIME mistakenly retrieves a document about financial margins (an off-topic false positive), PRF \ecl\ successfully identifies a document discussing geological active margins, which is the correct topic. 
By retrieving documents topically relevant to the query, \ecl\ increases the likelihood of positioning relevant documents higher in the ranking.

\renewcommand{\arraystretch}{1.2}
\begin{figure}[t]
\centering
\begin{tabular}{|>{\raggedright\arraybackslash}p{0.49\textwidth}|>{\raggedright\arraybackslash}p{0.49\textwidth}|}
\hline
\multicolumn{2}{|c|}{\textbf{Query:} ``What is an active margin''} \\
%enter table
\hline
\multicolumn{1}{|c|}{\textbf{PRF \ecl} (AP: 0.498)} & \multicolumn{1}{|c|}{\textbf{PRF Standard DIME} (AP: 0.183)} \\ 
\hline
\multicolumn{2}{|c|}{\textbf{Top Relevant Document:}} \\
\hline 
    % Left cell content with margin
    \hspace{0.1cm}\begin{minipage}[t]{0.46\textwidth}
    \textbf{Document [8446505]} An active \boldgreen{continental margin} refers to the submerged edge of a continent overriding an oceanic lithosphere \big[\ldots\big]. 
    \end{minipage} 
    &
    % Right cell content with margin
    \hspace{0.1cm}\begin{minipage}[t]{0.46\textwidth}
    \textbf{Document [6122722]} Best Answer: An \boldgreen{active margin is a tectonic plate boundary}, a passive margin is a compositional transition within a tectonic plate \big[\ldots\big]. \vspace{0.2cm}
\end{minipage} \\
\hline
\multicolumn{2}{|c|}{\textbf{First FPs documents:}} \\ 
\hline 
    % Left cell content with margin
    \hspace{0.1cm}\begin{minipage}[t]{0.46\textwidth}
    \textbf{Document [5286535]} \big[\ldots\big] \boldgreen{an active continental margin} is found on the leading edge of the continent where it is crashing into an oceanic platean \big[\ldots\big]. \hfill\break

    \textbf{Document [5286540]} The \boldgreen{continental margin} is the submerged outer edge of a continent. \big[\ldots\big].
    \end{minipage} 
    
    % Right nested column content with margin
    & \hspace{0.1cm}\begin{minipage}[t]{0.46\textwidth}
    \textbf{Document [5354863]} \boldred{The option margin is the cash} or securities an investor must deposit in his account as collateral before writing options. \big[\ldots\big]. \hfill\break
        
    \textbf{Document [6689166]} Answer Wiki. \boldred{Active Contingent} means that the seller has accepted an offer from a buyer, \big[\ldots\big].\vspace{0.2cm}
\end{minipage} \\
\hline
\end{tabular}
\caption{Comparison of retrieval results (Top Relevant document and First FPs documents) for the query "What is an active margin?" using PRF \ecl\ and PRF Standard DIME methods with the model ANCE. The green colour indicates if the documents match the same topic of the query and the red otherwise. PRF \ecl\ achieves a significantly higher AP. This improvement is attributed to \ecl's ability to push topically relevant documents to the query higher in the ranking.}
\end{figure}
\label{topic-separation}
\renewcommand{\arraystretch}{1.}

% experimental settings
\section{Experimental Setup}\label{exp-setting}

In our experiments, we compare our proposed \ecl\  against the state-of-the-art DIMEs for dense IR systems. We experiment with three dense retrieval models: ANCE \cite{ance}, Contriever \cite{izacard2021unsupervised}, and TAS-B \cite{tasb}, all of which have been fine-tuned using the MS MARCO \cite{bajaj2018msmarcohumangenerated} passage dataset. The pre-trained weights for these models are publicly available through the Huggingface platform, and the models operate within 768-dimensional embedding spaces. \hfill \break

\noindent
\textbf{Datasets.} Consistent with previous studies, we evaluate our methodology on three widely used benchmark collections for in-domain evaluation: TREC Deep Learning 2019 (DL '19) \cite{trec19}, TREC Deep Learning 2020 (DL '20) \cite{trec20}, and Deep Learning Hard (DL HD) \cite{trechd}. To assess the robustness and provide additional validation of our method, we further evaluate \ecl\ on out-of-domain data based on the TREC Robust '04 (RB '04)  collection~\cite{robust04}. Each in-domain query set focuses on ad-hoc passage retrieval and is composed of 43, 54, and 50 annotated queries, respectively, derived from the MS MARCO passage corpus. The RB ‘04 query set contains 249 queries and is based on the
TIPSTER disks 4 and 5, minus the congressional records corpus.\hfill \break

\noindent
\textbf{Metrics.} We evaluate the systems using standard metrics such as mean Average Precision (AP) and nDCG@10, following prior work and official benchmarks. To validate the effectiveness of our method to the baselines, we employed the paired Student’s $t$-test \cite{t-test}, after testing normality of distributions with Shapiro–Wilk test \cite{Shapiro} with a significance level of 0.05. The one-sided Wilcoxon \cite{wilcoxon} is chosen instead, as a non-parametric alternative for non-normal distributed data. Multiple hypothesis testing is performed with Holm-Bonferroni correction \cite{29def780-e117-38f0-8afb-edf384af3fad}.\hfill \break

\noindent
\textbf{Hyperparameters.}
We define four primary hyperparameters that influence different aspects of the model's decision-making process: $k^{+}$, $k^{-}$, $\alpha$, and $\beta$. The parameter $k^{-} \in \{2, \dots, 6 \}$ (resp. $k^{+} \in \{2, \dots, 14 \}$), determines the number of relevant (resp. irrelevant) documents, used to build our \textit{sun} (resp. \textit{moon}) embeddings. %
The hyperparameter $\alpha$ controls the strength of the relevant representative embedding while $\beta$ modulates the denoising effect of the irrelevant representative embedding. Both are positive values increasing linearly from 0.1 up to 1. \hfill \break

\noindent
\textbf{Baselines.} We compare our method to standard DIMEs: 
\begin{itemize}[noitemsep]
    \item PRF DIME: $u_q^{\text{PRF}}@1(i) = \bfq_i \cdot \bfp_i$, with  $\bfp$ the top relevant document for the query $q$.
    \item LLM DIME: $u_q^{\text{LLM}}(i) = \bfq_i \cdot \bfa_i$, with $\bfa$ being the embedding of the LLM generated answer to the query; we use GPT4 \cite{gpt4} as LLM in our experiments.
\end{itemize}
We set $k^{+}=1$ since it achieves the best results~\cite{DIME}. We will refer to the dense IR system at full dimensionality as \textit{Baseline}. \ecl\ uses a retrieved collection of documents $\mathcal{D}_q$ of size $1,000$.
 
%% results
\section{Results}\label{results}

In our experiments, we investigate the following research questions:
\begin{itemize}
    \item \textbf{RQ1}: Can pseudo-irrelevance feedback be leveraged to improve state-of-the-art DIME approaches?
    %\modificato{Does our proposed approach significantly improve over standard DIMEs?} 
    
    \item \textbf{RQ2}: What type of irrelevant documents are used to highlight relevant dimensions in the embedding space?
    %\modificato{Does our method actually depend on irrelevant documents to highlight relevant dimensions?}
    
    \item \textbf{RQ3}: Is our approach dependent on the semantic content of irrelevant documents to improve IR system performance?
    %\modificato{Is \ecl\ dependent on the semantic content of the irrelevant documents?} 
\end{itemize}

%\begin{itemize}
%    \item \textbf{RQ1}: \modificato{Does \ecl\ significantly improve  over standard DIMEs?}
%    \item \textbf{RQ2}: \modificato{Does \ecl\ actually depend on irrelevant documents to highlight relevant dimensions?}
%    \item \textbf{RQ3}: \modificato{Is \ecl\ dependent on the semantic content of the irrelevant documents?}
%\end{itemize}

\noindent
\textbf{Results for RQ1:} Table \ref{rq1-main-res} compare both versions of \ecl\ with standard DIMEs (PRF and LLM) and Baseline on the TREC DL '19, DL '20, DH, and RB '04 datasets, using the ANCE, Contriever, and TAS-B models. We report the best result among varying the percentage of retained dimensions. %da qui cominciato i risultati
\ecl\ exhibits superior performance in the traditional evaluation protocol, improving performance up to $19.50\%$ (resp. $22.35\%$) in AP and $11.42\%$ (resp. $13.10\%$) in nDCG@10 w.r.t. the DIME-based baseline (resp. the Baseline). In particular, both PRF \ecl\ and LLM \ecl\ show statistically significant improvement with respect to their DIME counterparts and Baseline. The bold represents the best result for each dataset and metric, having LLM \ecl\ the most effective approach.

To further highlight the efficiency of \ecl, we investigate the performance when projecting the embeddings at low dimensionality. Such a comparison helps us understand how well the models can maintain their performance when using fewer resources.   
In Table \ref{rq1-improvement} we report the percentage of improvement that LLM \ecl\  and PRF \ecl\, using \textit{half} of the embedding dimensions, achieve respectively to LLM DIME and PRF DIME. We test across different collections and bi-encoders.
%qui i risultati
The results demonstrate that \ecl\ consistently outperforms the DIMEs baseline, even with reduced dimensionality. For the ANCE model, \ecl\ shows substantial improvements, with statistically significant gains up to 42.01\% in AP and 18.15\% in nDCG@10. The Contriever model exhibits similar trends, with \ecl\  outperforming DIME across most datasets. The statistically significant improvements are up to 7.04\% in AP and 6.88\% in nDCG@10. The TAS-B model results further improve up to a 14.30\% increase in AP and 5.51\% on nDCG@10.

The results of our study provide strong evidence that \ecl\ statistically significantly improves over standard DIMEs across various datasets, models and evaluation metrics, particularly in its LLM-based version. The effectiveness of \ecl\ persisted even when using only 50\% of the original embedding dimensionality, demonstrating its efficiency in resource utilization.

\renewcommand{\arraystretch}{1.2}
\begin{table}[t]
\centering
\caption{\textbf{Comparison between \ecl\   and baselines.} Effectiveness metrics of \ecl\ and baselines on different query sets and bi-encoders. \textbf{Bold} shows the best result for each column, \underline{underlined} if LLM \ecl\  (resp. PRF \ecl) strictly outperform LLM DIME (resp. PRF DIME). Superscripts $\supa$ and $\supb$ indicate that the result is statistically significantly ($p < 0.05$) better than Baseline or standard DIMEs, respectively. }
%\scriptsize
%\hspace*{-0.7cm}
% total width number of columns
%\resizebox{!}{0.31\textwidth}{
\resizebox{1\textwidth}{!}{
\begin{tabular}{l @{\hskip 4pt} l @{\hskip 5pt}
                c @{\hskip 2pt} c @{\hskip 4pt}
                c @{\hskip 2pt} c @{\hskip 4pt}
                c @{\hskip 2pt} c @{\hskip 4pt}
                c @{\hskip 2pt} c }
    \toprule
    % columns under columns
    & &\multicolumn{2}{c}{\textbf{DL '19}} &\multicolumn{2}{c}{\textbf{DL '20}} &\multicolumn{2}{c}{\textbf{DL HD}} &\multicolumn{2}{c}{\textbf{RB '04}}\\
    \cmidrule(lr){3-4} \cmidrule(lr){5-6} \cmidrule(lr){7-8} \cmidrule(lr){9-10}
    Model & Filter & AP & nDCG@10 & AP & nDCG@10 & AP & nDCG@10 & AP & nDCG@10\\
\midrule
\multirow{4}{*}{ANCE} 
&Baseline (a)  &0.361  &0.645  &0.392  &0.646  &0.180  &0.334  &0.146  &0.384 \\ %\cline{2-10}
&PRF DIME (b)  & 0.370 & 0.657 & 0.392 & 0.649 & 0.184 & 0.340 & 0.150 & 0.386 \\
&LLM DIME (b)  & 0.370 & 0.663 & 0.397 & 0.658 & 0.186 & 0.346 & 0.149 & 0.397 \\ %\cline{2-10}
&PRF \ecl\ & \underline{0.406}\supa & \underline{0.669} & \underline{0.408}\supa & \underline{0.656}\supa & \underline{0.193}\supab & \underline{0.349} & \textbf{\underline{0.179}}\supab & \underline{0.402}\supab \\
&LLM \ecl\ &\textbf{\underline{0.424}}\supab &\textbf{\underline{0.706}}\supab &\textbf{\underline{0.411}}\supab &\textbf{\underline{0.665}} &\textbf{\underline{0.210}}\supa &\textbf{\underline{0.360}} &\underline{0.168}\supab &\textbf{\underline{0.430}}\supab\\
\midrule
\multirow{4}{*}{Contriever}
&Baseline (a) &0.494  &0.677  &0.478  &0.666  &0.241  &0.375  &0.239  &0.481 \\ %\cline{2-10}
&PRF DIME (b) & 0.509 & 0.692 & 0.496 &0.713  & 0.251 & 0.388 & 0.253 &0.486 \\
&LLM DIME (b) & 0.523 & 0.745 & 0.505 & 0.697 & 0.263 & 0.396 &0.263  &\textbf{0.526} \\ %\cline{2-10}
&PRF \ecl\ & \underline{0.542}\supa & \underline{0.710} & \underline{0.509}\supa & \textbf{\underline{0.720}}\supa & \textbf{\underline{0.275}}\supa & \textbf{\underline{0.410}}\supab & \textbf{\underline{0.267}}\supab & \underline{0.501}\supa \\
&LLM \ecl\ &\textbf{\underline{0.557}}\supab &\textbf{\underline{0.751}}\supa &\textbf{\underline{0.515}}\supab &\underline{0.706}\supa &\textbf{\underline{0.275}}\supa &\underline{0.397}\supa  &0.263\supa  &\textbf{0.526}\supa\\
\midrule
\multirow{4}{*}{TAS-B} 
&Baseline (a) &0.476  &0.719& 0.475&0.685&0.238&0.374&0.197&0.428\\ %\cline{2-10}
&PRF DIME (b) & 0.507 & 0.731 & 0.489 & 0.712 & 0.244 & 0.385 & 0.215 & 0.443 \\
&LLM DIME (b) & 0.527 & 0.768 & 0.496 & 0.705 & 0.265 & 0.408 & 0.218 & 0.471 \\ %\cline{2-10}
&PRF \ecl\ & \underline{0.550}\supab & \underline{0.745}\supa & \underline{0.509}\supab & \textbf{\underline{0.726}}\supab & \underline{0.278}\supab & \underline{0.420}\supab & \textbf{\underline{0.232}}\supab & \underline{0.460}\supa \\
&LLM \ecl\ &\textbf{\underline{0.557}}\supab &\textbf{\underline{0.775}}\supab &\textbf{\underline{0.511}}\supab  &\underline{0.711}\supa &\textbf{\underline{0.283}}\supab &\textbf{\underline{0.422}}\supab &\underline{0.222}\supa &\textbf{\underline{0.475}}\supa \\
\bottomrule
\end{tabular}}
\label{rq1-main-res}
\end{table}
\renewcommand{\arraystretch}{1.}

\renewcommand{\arraystretch}{1.2}
\begin{table}[t]
\centering
\caption{\textbf{Variation of metrics using half of the embedding dimensions.} Effectiveness metrics of \ecl\ and baselines on different query sets and bi-encoders using only 50\% of the original dimensionality. \sstar \ indicates a statistically significant result (paired $t$-test with $p < 0.05$).}
%\scriptsize
%\hspace*{-0.7cm}
% total width number of columns
%\resizebox{!}{0.31\textwidth}{
\resizebox{1\textwidth}{!}{
\begin{tabular}{l @{\hskip 4pt} l @{\hskip 5pt}
                c @{\hskip 2pt} c @{\hskip 4pt}
                c @{\hskip 2pt} c @{\hskip 4pt}
                c @{\hskip 2pt} c @{\hskip 4pt}
                c @{\hskip 2pt} c }
    \toprule
% columns under columns
& &\multicolumn{2}{c}{\textbf{DL '19}} &\multicolumn{2}{c}{\textbf{DL '20}} &\multicolumn{2}{c}{\textbf{DL HD}} &\multicolumn{2}{c}{\textbf{RB '04}}\\
\cmidrule(lr){3-4} \cmidrule(lr){5-6} \cmidrule(lr){7-8} \cmidrule(lr){9-10}
Model & Filter & AP & nDCG@10 & AP & nDCG@10 & AP & nDCG@10 & AP & nDCG@10\\
\midrule
\multirow[c]{6}{*}{ANCE} 
 & PRF DIME & 0.3156 & 0.6251 & 0.3382 & 0.6057 & 0.1568 & 0.3109 & 0.1133 & 0.3254 \\
 & PRF \ecl\ & 0.3751 & 0.6462 & 0.3766 & 0.6164 & 0.1707 & 0.3221 & 0.1609 & 0.3606 \\
 & \textbf{\%-improvement} & 18.85\sstar & 3.37\sstar & 11.35\sstar & 1.77 & 8.86\sstar & 3.60 & 42.01\sstar & 10.82\sstar \\ 
 \cmidrule{2-10}
& LLM DIME & 0.3238 & 0.6334 & 0.3427 & 0.5854 & 0.1566 & 0.3020 & 0.1157 & 0.3267 \\
 & LLM \ecl\ & 0.3949 & 0.6863 & 0.3725 & 0.6124 & 0.1962 & 0.3158 & 0.1401 & 0.3860 \\
 & \textbf{\%-improvement} & 21.96\sstar & 8.35\sstar & 8.69\sstar & 4.61\sstar & 25.29\sstar & 4.57 & 21.09\sstar & 18.15\sstar \\ 
 \midrule
\multirow[c]{6}{*}{Contriever} 

 & PRF DIME & 0.5051 & 0.6838 & 0.4959 & 0.7013 & 0.2486 & 0.3836 & 0.2507 & 0.4787 \\
 & PRF \ecl\ & 0.5405 & 0.7009 & 0.5057 & 0.7062 & 0.2654 & 0.4100 & 0.2669 & 0.4936 \\
 & \textbf{\%-improvement}  & 7.01\sstar & 2.50 & 1.98 & 0.70 & 6.76 & 6.88\sstar & 6.46\sstar & 3.11 \\ \cmidrule{2-10}
 &LLM DIME & 0.5207 & 0.7367 & 0.5007 & 0.6933 & 0.2570 & 0.3773 & 0.2622 & 0.5234 \\
 & LLM \ecl\ & 0.5572 & 0.7436 & 0.5104 & 0.6957 & 0.2751 & 0.3953 & 0.2630 & 0.5263 \\
 & \textbf{\%-improvement} & 7.01\sstar & 0.94 & 1.94\sstar & 0.35 & 7.04\sstar & 4.77 & 0.30 & 0.55 \\ 
 \midrule
\multirow[c]{6}{*}{TAS-B} 

 & PRF DIME & 0.5055 & 0.7278 & 0.4883 & 0.7103 & 0.2405 & 0.3849 & 0.2127 & 0.4344 \\
 & PRF \ecl\ & 0.5496 & 0.7397 & 0.5091 & 0.7199 & 0.2749 & 0.4061 & 0.2315 & 0.4545 \\
 & \textbf{\%-improvement} & 8.72\sstar & 1.63 & 4.26\sstar & 1.35 & 14.30\sstar & 5.51 & 8.84\sstar & 4.63\sstar \\ \cmidrule{2-10}
 & LLM DIME & 0.5227 & 0.7684 & 0.4962 & 0.6977 & 0.2607 & 0.3994 & 0.2173 & 0.4715 \\
 & LLM \ecl\ & 0.5547 & 0.7724 & 0.5069 & 0.6859 & 0.2791 & 0.3963 & 0.2122 & 0.4619 \\
 & \textbf{\%-improvement} & 6.12\sstar & 0.52 & 2.16 & -1.69 & 7.06\sstar & -0.78 & -2.35 & -2.04 \\ 
\bottomrule
\end{tabular}}
\label{rq1-improvement}
\end{table}
\renewcommand{\arraystretch}{1.}

\begin{figure}
\centering
\begin{subfigure}[b]{1\textwidth}
   \includegraphics[width=1\textwidth]{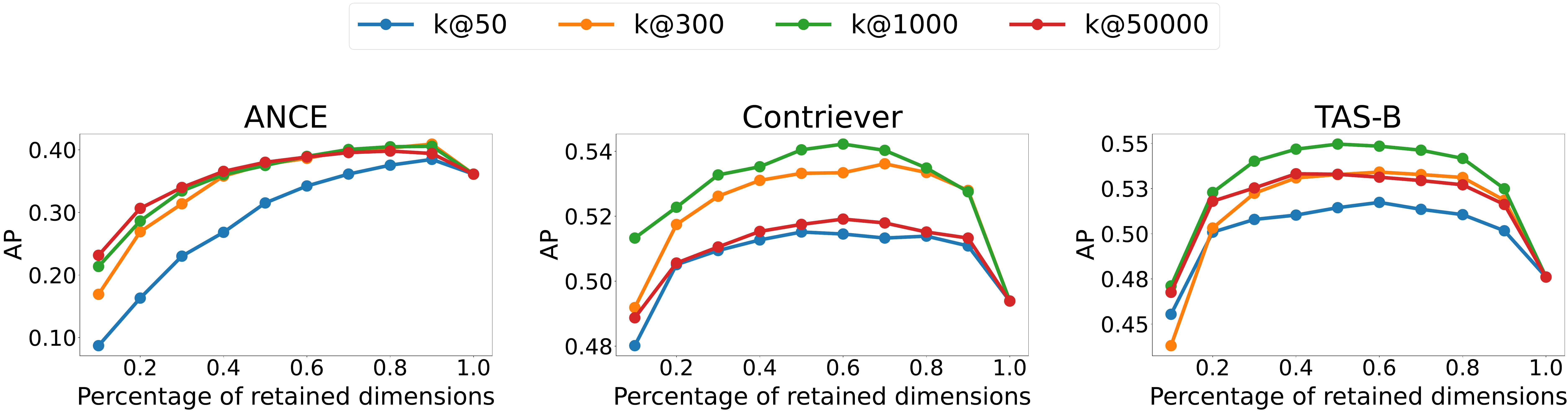}
   \caption{DL '19}
   \label{fig:dl19cutoff} 
\end{subfigure}
\vspace{0.2cm}
\begin{subfigure}[b]{1\textwidth}
   \includegraphics[width=1\textwidth]{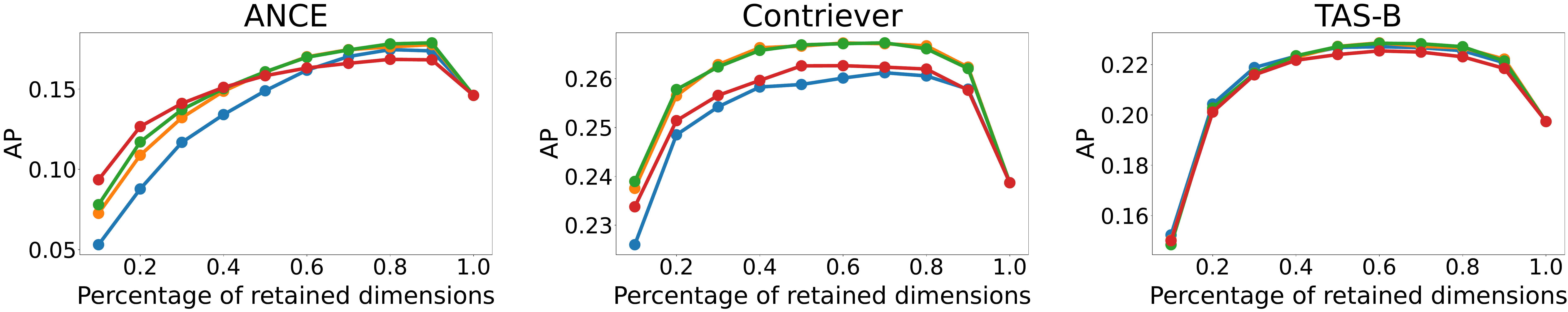}
   \caption{RB '04}
   \label{fig:dl20cutoff}
\end{subfigure}
\caption[caption]{Performance comparison of PRF-\ecl\ on DL '19 (a) and RB '04 (b) collections, showing AP as the percentage of retained dimensions increases. Different $k$ values represent the cardinality of retrieved document sets $\mathcal{D}_q$. Smaller $k$ (e.g. $k$=50) includes only highly relevant documents, while larger $k$ (e.g. $k=50,000$) gradually incorporates less relevant documents, affecting retrieval performance.}
\label{rq2-res}
\vspace{0.3cm}
\begin{subfigure}[b]{1\textwidth}
\includegraphics[width=1\textwidth]{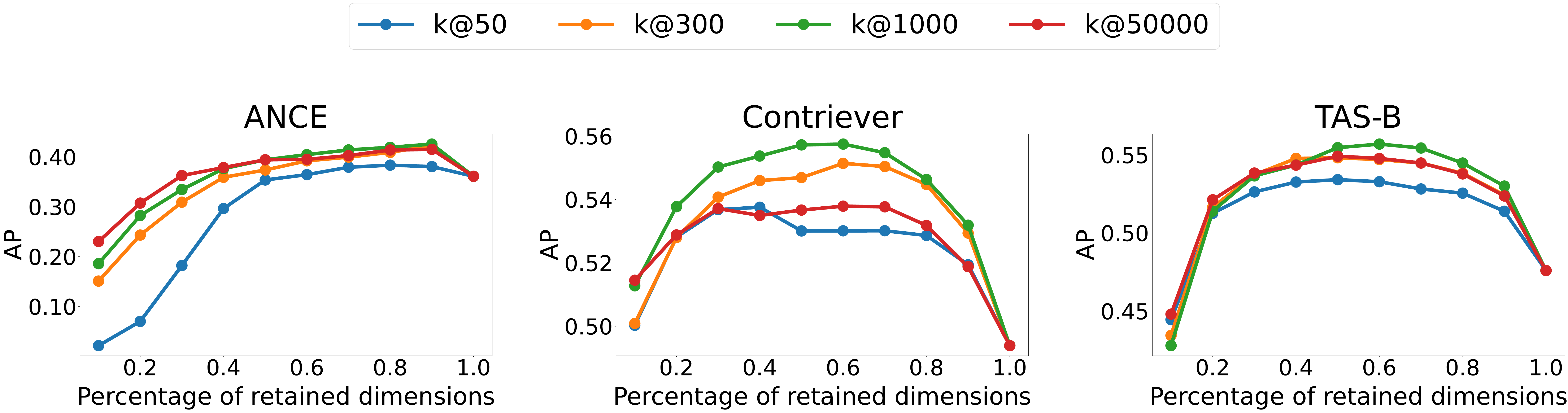}
   \caption{DL '19}
   \label{fig:dl19_gptcutoff} 
\end{subfigure}
\begin{subfigure}[b]{1\textwidth}
\includegraphics[width=1\textwidth]{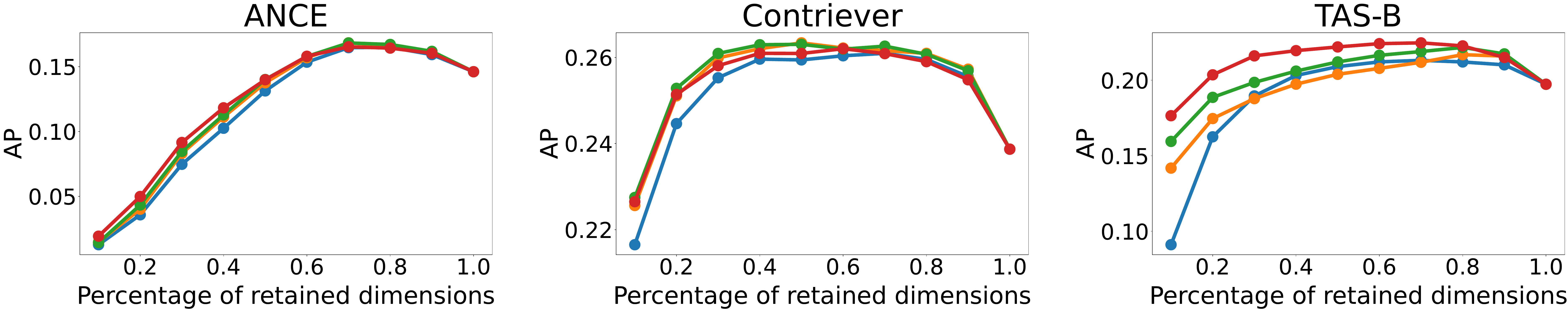}
   \caption{RB '04}
   \label{fig:dl20_gptcutoff}
\end{subfigure}
\caption[caption]{Performance comparison of LLM-\ecl\ on DL '19 (a) and RB '04 (b) collections, showing AP as the percentage of retained dimensions increases. Different $k$ values represent the cardinality of retrieved document sets $\mathcal{D}_q$. Smaller $k$ (e.g. $k$=50) includes only highly relevant documents, while larger $k$ (e.g. $k=50,000$) gradually incorporates less relevant documents, affecting retrieval performance.}
\label{rq2-res2}
\end{figure}

\hfill\break\noindent
\textbf{Results for RQ2:} We conducted an ablation study varying the cardinality $k$ of the collection $\mathcal{D}_q$ of retrieved documents at full dimensionality. We used PRF-based \ecl\ and measured performance using the AP metric as the percentage of retained dimension increased. The study included collections of 50, 300, 1000, and 50000 retrieved documents. Figures \ref{rq2-res} and \ref{rq2-res2} confirm that subtracting embeddings of documents with lower relevance scores yields improvement in retrieval performance. This suggests that using highly irrelevant documents (those not close to the pseudo-relevant ones) increases overall effectiveness. Notably, with $k$=50 (represented by the blue curves in Figures \ref{rq2-res} and \ref{rq2-res2}), where the bottom-${k^{-}}$ documents have relatively high relevance scores, the improvement is less pronounced. This indicates that using these documents to build the irrelevant representative vector is less effective, as it may lower both irrelevant and relevant dimensions.

To answer RQ2, our method favour the inclusion of highly irrelevant documents to effectively highlight relevant dimensions. As we increase $k$ and include documents with a lower relevance score, the method's performance improves, indicating its ability to leverage the contrast between relevant and irrelevant information.

\renewcommand{\arraystretch}{1.2}
\begin{table}
\centering
\caption{\textbf{Random sampling analysis of irrelevant documents.} Retrieval performance comparison of PRF-based \ecl\ on different query sets and bi-encoders. The methods operate on a fixed collection of top-1000 documents retrieved by each base model to build the irrelevant representation vector. Three methods randomly sample $k^{-}$ documents from the bottom-30, 100, and 150 of the collection, respectively, while the fourth serves as a baseline using a fixed bottom-$k^{-}$ without sampling.}
%\scriptsize
\resizebox{1.0\textwidth}{!}{
\small
% total width number of columns
\begin{tabular}{l @{\extracolsep{5pt}} l @{\hskip 10pt} c @{\hskip 5pt} c @{\hskip 15pt}c @{\hskip 5pt}c}
\toprule
% columns under columns
& &\multicolumn{2}{c}{\textbf{DL '19}} &\multicolumn{2}{c}{\textbf{DL '20}}\\
\cmidrule(lr){3-4} \cmidrule(lr){5-6}
Model & Sampling & AP & nDCG@10 & AP & nDCG@10\\
\midrule
\multirow{4}{*}{ANCE} 
&$@30$  & 0.4049 ± 0.0038  & 0.6550 ± 0.0021  & 0.4065 ± 0.0015 & 0.6533 ± 0.0016 \\
&$@100$ & 0.4051 ± 0.0042  & 0.6559 ± 0.0035  & 0.4047 ± 0.0033 & 0.6530 ± 0.0026 \\
&$@150$ & 0.4054 ± 0.0067  & 0.6540 ± 0.0024  & 0.4038 ± 0.0036 & 0.6530 ± 0.0010 \\
&Bottom-${k^{-}}$ &0.406&0.662&0.408&0.656\\
\midrule
\multirow{4}{*}{Contriever}
&$@30$  & 0.5351 ± 0.0014  & 0.7032 ± 0.0032  & 0.5038 ± 0.0016  & 0.7022 ± 0.0013 \\
&$@100$ & 0.5333 ± 0.0051  & 0.7026 ± 0.0047  & 0.5047 ± 0.0016  & 0.6961 ± 0.0031 \\
&$@150$ & 0.5334 ± 0.0021  & 0.7037 ± 0.0016  & 0.5058 ± 0.0021  & 0.6964 ± 0.0026 \\
&Bottom-${k^{-}}$ &0.542&0.710&0.509&0.708\\
\midrule
\multirow{4}{*}{TAS-B} 
&$@30$  & 0.5454 ± 0.0029  & 0.7404 ± 0.0033 & 0.5057 ± 0.0034 & 0.7168 ± 0.0042 \\
&$@100$ & 0.5492 ± 0.0033  & 0.7401 ± 0.0032 & 0.5050 ± 0.0030 & 0.7194 ± 0.006 \\
&$@150$ & 0.5522 ± 0.0035  & 0.7400 ± 0.0043 & 0.5069 ± 0.0038 & 0.7186 ± 0.0014 \\
&Bottom-${k^{-}}$ &0.550&0.745&0.509&0.726\\
\bottomrule
\end{tabular}
}
\resizebox{1\textwidth}{!}{
\begin{tabular}{l @{\extracolsep{5pt}} l @{\hskip 10pt} c @{\hskip 5pt} c @{\hskip 15pt}c @{\hskip 5pt}c}
% columns under columns
& &\multicolumn{2}{c}{\textbf{DL HD}} &\multicolumn{2}{c}{\textbf{RB '04}}\\
\cmidrule(lr){3-4} \cmidrule(lr){5-6}
Model & Sampling & AP & nDCG@10 & AP & nDCG@10\\
\midrule
\multirow{4}{*}{ANCE} 
&$@30$  & 0.1901 ± 0.0017  & 0.3376 ± 0.0031  & 0.1782 ± 0.0012 & 0.3999 ± 0.0020 \\
&$@100$ & 0.1913 ± 0.0012  & 0.3374 ± 0.0023  & 0.1790 ± 0.0008 & 0.4002 ± 0.0023 \\
&$@150$ & 0.1903 ± 0.0005  & 0.3367 ± 0.0029  & 0.1795 ± 0.0008 & 0.4020 ± 0.0026 \\
&Bottom-${k^{-}}$ &0.193&0.343&0.179&0.401\\
\midrule
\multirow{4}{*}{Contriever}
&$@30$  & 0.2714 ± 0.0039  & 0.4092 ± 0.0035  & 0.2679 ± 0.0008  & 0.4981 ± 0.0013 \\
&$@100$ & 0.2696 ± 0.0030  & 0.4054 ± 0.0026  & 0.2680 ± 0.0006  & 0.4999 ± 0.0012 \\
&$@150$ & 0.2743 ± 0.0010  & 0.4031 ± 0.0017  & 0.2676 ± 0.0011  & 0.5000 ± 0.0024 \\
&Bottom-${k^{-}}$ &0.275&0.410&0.267&0.501\\
\midrule
\multirow{4}{*}{TAS-B} 
&$@30$  & 0.2755 ± 0.0022  & 0.4168 ± 0.0047  & 0.2253 ± 0.0001  & 0.4567 ± 0.0017 \\
&$@100$ & 0.2719 ± 0.0031  & 0.4164 ± 0.0053  & 0.2255 ± 0.0001  & 0.4569 ± 0.0030 \\
&$@150$ & 0.2757 ± 0.0042  & 0.4160 ± 0.0047  & 0.2255 ± 0.0003  & 0.4571 ± 0.0024 \\
&Bottom-${k^{-}}$ &0.278&0.420&0.232&0.460\\
\bottomrule
\end{tabular}
}
\label{rq3-res}
\end{table}
\renewcommand{\arraystretch}{1.}

%\subsection{Robust Random Sampling Analysis: Full Corpus vs. Cutoff Points}
\hfill\break\noindent
\textbf{Results for RQ3:} Our previous experiments have demonstrated that \ecl\ is an effective DIME and that irrelevant documents play a crucial role in identifying relevant dimensions. In this final research question, we investigate whether the semantic content of these irrelevant documents significantly impacts performance. We designed our experiment as follows: (1) We fixed the collection $\mathcal{D}_q$, with $|\mathcal{D}_q| = 1000$; (2) Instead of selecting the exact bottom-$k^{-}$ documents, we randomly sampled $k^{-}$ documents from the last 30, 100, and 150 documents of that collection. We conducted this experiment using PRF-based \ecl\ and evaluated performance using AP and nDCG@10 metrics across benchmark collections and models. Due to the random sampling, we report mean performance scores and variances across multiple runs. Table \ref{rq3-res} presents our findings. The results indicate no significant difference between randomly sampling irrelevant documents from the lower end of the ranked list and selecting the exact bottom-$k^{-}$ documents.

To conclude, answering our RQs, leads us to conclude that the effectiveness of irrelevant documents in \ecl\ stems primarily from their low relevance scores, rather than from any specific semantic properties of the text itself.

% conclusions
\section{Conclusion}\label{conclusion}

We introduced \ecl, a novel approach that enhances dense retrieval by uniquely leveraging pseudo-irrelevant feedback to better distinguish between relevant and non-relevant dimensions in document embeddings. This use of irrelevant documents as a contrastive signal marks a significant shift from traditional DIME methods focused solely on relevant data.
Our extensive experiments revealed that to discover relevant dimensions, researchers should consider using documents with lower relevance scores as determined by the model, regardless of the text's semantic content. This approach proved effective in identifying and eliminating less informative dimensions.
We validate our method using four different datasets and three different models showing that \ecl\ achieve an average improvement of up to $19.50\%$ (resp. $22.35\%$) in AP and $11.42\%$ (resp. $13.10\%$) in nDCG@10 w.r.t. the DIME-based baseline (resp. the baseline using all dimensions). 
\ecl\ effectively highlights relevant dimensions, pushing documents with moderate relevance scores higher in the ranking. This results in a notable increase in AP. However, to achieve substantial improvements also in nDCG, future research should explore methods to identify the most relevant dimensions that can locate on the top ranking position the highest relevant documents.

% biblio
\bibliographystyle{splncs04}
\bibliography{bib}

\end{document}